\documentclass[12pt]{article} 
\usepackage{epsfig, amsmath, amssymb}
\usepackage{graphicx,psfrag} 
\newcommand{\be}{\begin{equation}}
\newcommand{\ee}{\end{equation}}
\newcommand{\ben}{\begin{equation}}
\newcommand{\een}{\end{equation}}
\newcommand{\bea}{\begin{eqnarray}}
\newcommand{\eea}{\end{eqnarray}}
\newcommand{\bA}{\begin{array}}
\newcommand{\eA}{\end{array}}
\newcommand{\bc}{\begin{center}}
\newcommand{\ec}{\end{center}}
\newcommand{\al}{\alpha}

\newcommand{\ra}{\rightarrow}
\newcommand{\del}{\partial}

\newcommand{\ie}{{\it i.e.}}
\newcommand{\eg}{{\it e.g.}}

\newcommand{\vx}{{\vec x}}

\newcommand{\cO}{{\cal O}}
\newcommand{\cI}{{\cal I}}

\begin{document}


\begin{titlepage}

\bc

\hfill 
\\         [40mm]

{\Huge de Sitter extremal surfaces}
\vspace{16mm}

{\large  K.~Narayan} \\
\vspace{3mm}
{\small \it Chennai Mathematical Institute, \\}
{\small \it SIPCOT IT Park, Siruseri 603103, India.\\}

\ec
\medskip
\vspace{40mm}

\begin{abstract}
We study extremal surfaces in de Sitter space in the Poincare slicing
in the upper patch, anchored on spatial subregions at the future
boundary ${\cal I}^+$, restricted to constant boundary Euclidean time
slices (focussing on strip subregions). We find real extremal surfaces
of minimal area as the boundaries of past lightcone wedges of the
subregions in question: these are null surfaces with vanishing
area. We also find complex extremal surfaces as complex extrema of the
area functional, and the area is not always real-valued. In $dS_4$ the
area is real. The area has structural resemblance with entanglement
entropy in a dual $CFT$. There are parallels with analytic
continuation from the Ryu-Takayanagi expressions for holographic
entanglement entropy in $AdS$.  We also discuss extremal surfaces in
the $dS$ black brane and the de Sitter ``bluewall'' studied
previously.  The $dS_4$ black brane complex surfaces exhibit a real
finite cutoff-independent extensive piece. In the bluewall geometry,
there are real surfaces that go from one asymptotic universe to the
other through the Cauchy horizons.
\end{abstract}

\end{titlepage}

{\tiny 
\begin{tableofcontents}
\end{tableofcontents}
}


\section{Introduction}

de Sitter space is fascinating for many reasons, in particular for
holographic explorations towards addressing questions of cosmology and
time. In this regard, some versions of $dS/CFT$ duality
\cite{Strominger:2001pn,Witten:2001kn,Maldacena:2002vr} associate to
de Sitter space a dual Euclidean CFT on the future timelike infinity
${\cal I}^+$ boundary\ (in the Poincare slicing).  A concrete
realization in the context of higher spin theories appears in
\cite{Anninos:2011ui}. Further work on $dS/CFT$ appears in \eg\
\cite{Harlow:2011ke,Ng:2012xp,Anninos:2012qw,Das:2012dt,Anninos:2012ft,
Anninos:2013rza,Das:2013qea,Banerjee:2013mca,Das:2013mfa}.

In $AdS/CFT$, there has been considerable interest in understanding
information theoretic notions in terms of geometric quantities via
holography, in particular stemming from the Ryu-Takayanagi
prescription \cite{Ryu:2006bv,Ryu:2006ef} (see
\cite{HEEreview,HEEreview2} for reviews) for calculating holographic
entanglement entropy of a subsystem in the strongly coupled boundary
field theory. This is the area of a bulk minimal surface (in Planck
units) anchored at the subsystem interface and dipping inwards upto a
certain maximal depth typically called the turning point. A different
way to think about this appears in \cite{Lewkowycz:2013nqa}.  More
generally, these are extremal surfaces \cite{HRT}. In this light, one
might speculate that the bulk subregion enclosed by the entangling
surface and the boundary subsystem in some sense encodes bulk physics
corresponding to that part of the boundary theory contained in the
subsystem, although a detailed understanding of the hologram (and bulk
locality) would seem more intricate.

It is interesting to consider these questions in the context of de
Sitter space and $dS/CFT$. Assuming there is translation invariance
with respect to a boundary Euclidean time direction, imagine
constructing a subregion on a Euclidean time slice of the future
boundary ${\cal I}^+$.  Tracing out the complement of this subregion
would lead to some loss of information and thereby give some
associated entropy, which one might attribute to the subregion being
entangled with the complement. In the bulk, intuition from the
Ryu-Takayanagi prescription in $AdS/CFT$ suggests that we study
extremal surfaces in de Sitter space (in the Poincare slicing) on a
constant boundary Euclidean time slice, defined as anchored on the
subregion on the future (spacelike) boundary and dipping inwards (\ie\
in the bulk time direction, towards the past).  We find (sec.~2) that
the bulk extremization problem exhibits some crucial sign differences
from the $AdS$ case. Focussing first on real surfaces, there are
correspondingly some technical differences such as the absence of a
natural turning point (where the surface stops dipping inward). For
sufficiently symmetric subregions such as strips (with an axis of
symmetry), extremal surfaces can be defined as the union of two
half-extremal-surfaces joined continuously but with a sharp cusp. Upon
requiring that we choose minimal area, the extremal surfaces become
null surfaces with zero area. In fact these are simply the boundaries
of the past lightcone wedges of the subregion in question (restricted 
to the boundary Euclidean time slice), and are thus analogous to the 
causal wedges associated with causal holographic information 
\cite{Hubeny:2012wa}\ (note that these bulk causal wedges and the 
corresponding causal holographic information in general do not
coincide with the bulk entangling subregion, and entanglement
entropy). This answer -- restrictions of past lightcone wedges -- 
is well-defined for arbitrary boundary subregions, even without 
sufficiently high symmetry, and gives vanishing area. These surfaces 
with vanishing area do not appear to have any connection to 
entanglement in $dS/CFT$.

It is therefore interesting to look for other extrema, in particular
complex saddle points of the extremization problem, motivated by
considerations in $dS/CFT$. For instance, in the formulation of
\cite{Maldacena:2002vr} of the $dS/CFT$ dictionary, the CFT partition
function is $Z_{CFT}=\Psi$ where $\Psi$ is the bulk late time
Hartle-Hawking wavefunction of the universe subject to appropriate
(Bunch-Davies) boundary conditions at early times. In a semiclassical
approximation $\Psi\sim e^{iS_{cl}}$, the dual CFT energy-momentum
tensor $\langle TT \rangle$ correlators exhibit central charge
coefficients of the form ${\cal C}_d\sim i^{1-d} {R_{dS}^{d-1}\over
  G_{d+1}}$\ (which is essentially an analytic continuation from
Euclidean $AdS_{d+1}$).  With this in mind, focussing again on strip
subregions in the present context, we indeed find these complex
extremal surfaces: they exhibit ``turning points'' in the
interior. They should be thought of as living in some auxiliary space,
and are distinct from the bulk past lightcone wedges (which define
real subregions in bulk $dS_4$). The area of these surfaces is in
general not real-valued. In $dS_4$, we find that $x(\tau)$
parametrizing the strip width being real-valued suggests that the bulk
time $\tau$ parametrizes a complex path $\tau=iT$. The area (in Planck
units) of these complex surfaces in $dS_4$ is real-valued and
negative, while in $dS_{d+1}$ with $d$ even, the nature of these
surfaces is different and the area is pure imaginary.  The area has
structural resemblance with entanglement entropy in a dual
(non-unitary) CFT$_d$, with central charge\ ${\cal C}_d\sim i^{1-d}
{R_{dS}^{d-1}\over G_{d+1}}$~, with a leading area law divergence, and
subleading terms.  There are parallels with analytic continuation from
the Ryu-Takayanagi holographic entanglement expressions from $AdS$.
It is a useful consistency check that these central charges\ ${\cal
  C}_d\sim i^{1-d} {R_{dS}^{d-1}\over G_{d+1}}$ here resemble those in
the $\langle TT \rangle$ correlators in \cite{Maldacena:2002vr},
mentioned above. From the point of view of the dual Euclidean CFT
living on the future boundary ${\cal I}^+$, one might formally
associate a density matrix w.r.t. boundary Euclidean time evolution
and a reduced density matrix to the subregion obtained by tracing out
the complement. It would be interesting to explore this further,
perhaps in $dS/CFT$ as entanglement entropy in the dual Euclidean CFT.

We then discuss (sec.~3) an asymptotically de Sitter space
\cite{Das:2013mfa} -- the $dS$ black brane -- where subleading
normalizable metric components are turned on: in $dS_4/CFT_3$, they
have the interpretation of saddle points representing the Euclidean
CFT with uniform energy-momentum density expectation value. The
corresponding extremal surfaces in the $dS_4$ black brane exhibit a
finite cutoff-independent real-valued extensive piece (again negative)
with some resemblance to a thermal entropy. Finally we discuss (real)
extremal surfaces in the closely related $dS$ ``bluewall'' geometry,
which are not obtained by analytic continuation: there are real
extremal surfaces which cross from one asymptotic universe to the
other through the Cauchy horizons.

\section{Extremal surfaces in de Sitter space}

de Sitter space $dS_{d+1}$ in the Poincare slicing or planar coordinate 
foliation is given by the metric
\be\label{dSpoinc}
ds^2 = {R_{dS}^2\over\tau^2} (-d\tau^2 + dw^2 + dx_i^2)\ ,
\ee
where half of the spacetime, \eg\ the upper patch, has ${\cal I}^+$ 
at $\tau = 0$ and a coordinate horizon at $\tau=-\infty$. This may be 
obtained by analytic continuation of a Poincare slicing of $AdS$, 
\be\label{AdStodS}
r \rightarrow -i\tau\ ,\qquad R_{AdS} \rightarrow -iR_{dS}\ ,\qquad 
t\ra -iw\ ,
\ee
where $w$ is akin to boundary Euclidean time, continued from time in $AdS$.

The dual Euclidean CFT is taken as living on the future $\tau=0$
boundary ${\cal I}^+$.  We assume translation invariance with respect
to a boundary Euclidean time direction, say $w$, and consider a
subregion on a $w=const$ slice of ${\cal I}^+$. One might imagine that
tracing out the complement of this subregion then gives entropy in
some sense stemming from the information lost.  In the bulk, we study
de Sitter extremal surfaces on the $w=const$ slice, analogous to the
Ryu-Takayanagi prescription in $AdS/CFT$. Operationally these extremal
surfaces begin at the interface of the subsystem (or subregion) and
dip inwards (towards the past, in the bulk time direction).  For
simplicity, consider a strip on the $w=const$ surface (\ie\ a constant
boundary Euclidean time surface): this bulk $d$-dim subspace has
metric
\be\label{dSPoinc-wslice}
ds^2 = {R_{dS}^2\over\tau^2} \Big(-d\tau^2 + \sum_{x_i\neq w} dx_i^2\Big) .
\ee
This is not a spacelike subspace in the bulk and it might seem that 
the extremal surfaces are timelike in general: however we will find 
that this is not the case.

\subsection{Real extremal surfaces}

\begin{figure}[h] 
\bc \includegraphics[width=25pc]{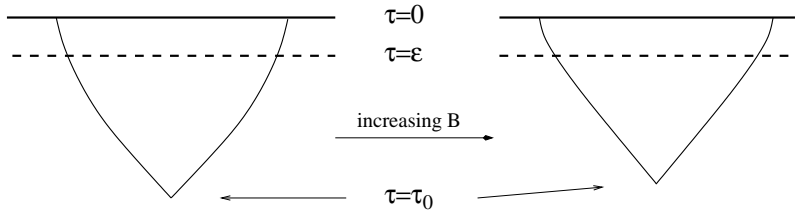} 
\caption{{\label{eedS}\footnotesize {Extremal surfaces in de Sitter 
made of two half-extremal surfaces joined continuously but with a 
sharp cusp at $\tau_0$.\ As $B$ increases (till eventually 
$B\gg {1\over\epsilon^{d-1}}$), the surface approaches ${\dot x}^2\ra 1$ 
(figure on the right).}}}
\ec
\end{figure}

Let us consider a strip subregion with width direction say $x$, the 
remaining $x_i$ being labelled $y_i$. A bulk surface on the $w=const$ 
slice bounding this subregion and dipping inward (towards the past) is 
bulk codim-2: its area functional in Planck units is
\be\label{RTdS01}
S_{dS} = {1\over 4G_{d+1}} 
\int \prod_{i=1}^{d-2} {R_{dS}dy_i\over\tau} {R_{dS}\over\tau} 
\sqrt{d\tau^2 - dx^2} = {R_{dS}^{d-1} V_{d-2}\over 4G_{d+1}} 
\int {d\tau\over\tau^{d-1}} \sqrt{1- \Big({dx\over d\tau}\Big)^2}\ .
\ee
We consider extremizing the action to find extremal surfaces with 
minimal area, along the lines of the Ryu-Takayanagi prescription for 
entanglement entropy in $AdS$. The $S_{dS}$ extremization gives a 
conserved quantity\ (${\dot x}\equiv {dx\over d\tau}$)
\be\label{RTdS02}
-{{\dot x}\over\sqrt{1-{\dot x}^2}} = B\tau^{d-1}\ 
\quad \Rightarrow\quad 1-{\dot x}^2 = {1\over 1+B^2\tau^{2d-2}}
\quad\ \ie\ \qquad {\dot x}^2 = {B^2\tau^{2d-2}\over  1+B^2\tau^{2d-2}}\ .
\ee
We see that ${\dot x}^2\ra 0$ near the boundary $\tau\ra 0$.
Assuming the conserved constant satisfies $B^2>0$ makes all the 
expressions real-valued and means ${\dot x}^2>0$, with 
${\dot x}^2\ra 1$ in the deep interior for large $|\tau|$. For 
$B^2>0$, these are timelike surfaces\footnote{One might instead 
want to consider spacelike surfaces with 
${\dot x}^2>1$ and therefore take, instead of (\ref{RTdS01}), the 
area functional as\ $S_{dS} = {R_{dS}^{d-1} V_{d-2}\over 4G_{d+1}} 
\int {d\tau\over\tau^{d-1}} \sqrt{({dx\over d\tau})^2-1}$. We will 
discuss this in the next subsection.}. This gives the solution (upto 
boundary conditions) and corresponding area integral
\be\label{RTdS}
x(\tau) = \pm \int {B\tau^{d-1} d\tau\over\sqrt{1+B^2\tau^{2d-2}}} 
\equiv \pm X(\tau)\ ,\qquad
S_{dS} = {R_{dS}^{d-1}\over 4G_{d+1}} V_{d-2} \int_\epsilon^{\tau_0}
{d\tau\over\tau^{d-1}} {1\over\sqrt{1+B^2\tau^{2d-2}}}\ .
\ee
The main difference between this case and the minimal surface in $AdS$ 
stems from $B^2>0$ implying that there is no smooth ``turning point'' 
where ${\dot x}^2 = {B^2\tau^{2d-2}\over 1+B^2\tau^{2d-2}} \ra\infty$. 
In fact $B^2>0$ means ${\dot x}^2$ is bounded, with $0\leq {\dot x}^2\leq 1$.
For any finite $B^2>0$, the extremal surface in this case begins to 
dip inwards from one boundary of the strip subregion and (rather than 
turning around as in $AdS$) continues indefinitely, eventually 
approaching ${\dot x}\ra \pm 1$.\ With a view to associating a bulk 
subregion with the boundary subregion in question, let us artifically 
cut off the inward dipping surface at some interior location 
$\tau=\tau_0$, the bulk subregion then defined by the interior of 
the boundary strip subregion and the joined surface. So consider 
the half-extremal-surfaces,
\bea
&& \qquad\qquad x_L(\tau)=X(\tau)-X(\tau_0)=-x_R(\tau)\ , \nonumber\\
&& x_L(\tau_0)=0=x_R(\tau_0) ,\quad\ x_L(0)=-{l\over 2} = -x_R(0)\qquad 
\Rightarrow\quad   {l\over 2} = X(\tau_0)\ .
\eea
This gives an extremal surface made of two half-extremal-surfaces
joined continuously but with a sharp cusp at $\tau_0$\ (see
Figure~\ref{eedS}). This defines the corresponding wedge-like bulk
subregion, enclosed by this extremal surface and the boundary
subregion. These conditions do not determine the parameters $B,\tau_0$
uniquely, given the subregion width $l$.  Varying $B$ gives different
extremal surfaces. By comparison, in the $AdS$ case, the turning point
$\tau_*={1\over B}$ is fixed by the global nature of the entangling
surface as the location where ${\dot x}^2\ra\infty$, the surface
turning around.

To follow the Ryu-Takayanagi prescription, we would want to identify
those extremal surfaces that have minimal area\footnote{Note that 
surfaces with maximum area correspond to minimizing $B$:\ this gives 
$B=0$, which are disconnected surfaces $x(\tau)=const$, with area
$S_{dS}\sim {R_{dS}^{d-1}\over 4G_{d+1}} V_{d-2} \int_\epsilon^{\tau_0}
{d\tau\over\tau^{d-1}}$ with a leading divergent piece.}. From
(\ref{RTdS}), we see that as $B$ increases, the area $S_{dS}$
decreases. Furthermore, (\ref{RTdS02}) shows that as $B$ increases,
${\dot x}^2$ increases and eventually approaches ${\dot x}^2\ra 1$ as
$B\ra\infty$.  In this limit, $x(\tau)\ra\pm\tau$ and $S_{dS}$ appears
to vanish. In fact this is a sensible result: in hindsight, it should
have been obvious from (\ref{RTdS01}) that minimal area arises when
the extremal surface becomes null. This null extremal surface is in
fact simply the boundary of the past bulk lightcone of the subregion, 
restricted to the boundary Euclidean time slice.

An alternative argument corroborating the above conclusion is the 
following. Physically, the shortest length (or time) scale here is\ 
$\tau_{UV}=\epsilon$\ so that in (\ref{RTdS02}) when $B\epsilon^{d-1}\gg 1$ 
we can approximate ${\dot x}^2\sim 1$ and so $x(\tau)\sim \pm\tau$ 
giving ${l\over 2}\sim \tau_0$.\ Thus one might estimate (\ref{RTdS}) as
\be
S_{dS} \sim {R_{dS}^{d-1}\over 4G_{d+1}} V_{d-2} \int_\epsilon^{\tau_0}
{d\tau\over\tau^{d-1}}\ +\ {R_{dS}^{d-1}\over 4G_{d+1}} V_{d-2} 
\int_0^{\tau_0} {d\tau\over\tau^{d-1}} 
\Big({1\over\sqrt{1+B^2\tau^{2d-2}}} - 1\Big),
\ee
where the second integral can be seen to vanish as $\tau\ra 0$.
Now the first integral scales as ${V_{d-2}\over\epsilon^{d-2}}$ while 
the second integral can be expressed in terms of the hypergeometric 
function ${}_2F_1$ as\ 
${R_{dS}^{d-1}\over G_{d+1}} {V_{d-2}\over (d-2) \tau_0^{d-2}} [1-
{}_2F_1({1\over 2}, -{d-2\over 2(d-1)}, {d\over2(d-1)}, -B^2\tau_0^{2d-2})]$\ 
\ (the extremal surface in (\ref{RTdS}) is itself expressed as\
$x(\tau) = \pm \tau~ {}_2F_1({1\over 2} , {1\over 2-2d} , 
{3-2d\over 2-2d} , -{1\over B^2\tau^{2d-2}})$\ or
$\pm B\tau^d~ {}_2F_1({1\over 2} , {d\over 2d-2} , {3d-2\over 2d-2} , 
-B^2\tau^{2d-2})$, using the integral representations of ${}_2F_1$). 
As $B^2$ increases, this second integral is seen to scale as 
$-{R_{dS}^{d-1}\over G_{d+1}} V_{d-2} B^{(d-2)/(d-1)}$.\ 
Thus when $B\sim {1\over\epsilon^{d-1}}$ this cancels the earlier 
contribution and we again see the leading $S_{dS}$ scaling to be vanishing.

In $dS_3$ (\ie\ $d=2$), we obtain\ 
$x(\tau) = \pm \int {B\tau d\tau\over \sqrt{1+B^2\tau^{2}}} 
= \pm {1\over B} \sqrt{1+B^2\tau^2}$ and the boundary conditions give\
${1\over B}(\sqrt{1+B^2\tau_0^2}-1) = {l\over 2}$, the area integral 
becoming\ $S_{dS} = {R_{dS}\over 4G_3} \int_\epsilon^{\tau_0}
{d\tau\over\tau} {1\over\sqrt{1+B^2\tau^{2}}}$. Analysing these 
vindicates the conclusions above.

It is worth noting that our construction of joining two
half-extremal-surfaces appears invalid unless the subsystem has
sufficiently high symmetry (in particular an axis of symmetry).
Relatedly, one might look askance at the entire extremization
procedure here, in particular whether one allows non-smooth surfaces
(with cusps) in the extremization: one might then wonder if more
general surfaces need to be considered, \eg\ a zigzag null surface
formed by joining multiple partial surfaces with multiple cusps. This
would be useful to systematise more rigorously.  However the final
answer, the past lightcone wedge, is well-defined for an arbitrary
subregion, comprising two piecewise smooth extremal surfaces joined
with just a single cusp (rather than multiple cusps). The past
lightcone wedge boundary (restricted to the boundary Euclidean time
slice) is however a complicated surface: it would be interesting to
understand the shape dependence here. The resulting area is of course
always zero for all these null surfaces, and does not reflect
entanglement structure.

From the point of view of bulk de Sitter alone, one could consider
volume subregions in the full $d$-dim boundary ${\cal I}^+$ (at
$\tau=0$), \ie\ not on the constant boundary Euclidean time
slice. These would give codim-1 surfaces. For a strip subregion with
width direction say $x$, the remaining $x_i$ being labelled $y_i$,
analysing the area integral of the bulk surface for extremization
gives
\bea\label{RTdS-codim1-1}
&& S_{dS} \sim\ R_{dS}^{d} V_{d-1}
\int {d\tau\over\tau^{d}} \sqrt{1- \Big({dx\over d\tau}\Big)^2}\ ,\qquad
-{{\dot x}\over\sqrt{1-{\dot x}^2}} = B\tau^{d}\ ,\nonumber\\
&& \Rightarrow\quad 
{\dot x}^2 = {B^2\tau^{2d}\over  1+B^2\tau^{2d}}\ ,\qquad
S_{dS} = R_{dS}^{d} V_{d-1} \int_\epsilon^{\tau_0}
{d\tau\over\tau^{d}} {1\over\sqrt{1+B^2\tau^{2d}}}\ .\quad
\eea
This has volume scaling. 
Again $S_{dS}$ decreases with increasing $B$, with ${\dot x}^2\ra 1$: 
the resulting extremal surfaces are null surfaces defining the past 
lightcone wedges of the volume subregion, with vanishing area.

These real null surfaces with vanishing area do not appear to 
have any bearing on entanglement in $dS/CFT$. In what follows, we 
will explore other complex saddle points.

\subsection{Complex extremal surfaces}

For what follows, it is useful to recall the $dS/CFT$ correspondence,
for de Sitter space in the Poincare slicing (\ref{dSpoinc}), obtained
by analytic continuation (\ref{AdStodS}) of Poincare $AdS$. A version
of $dS/CFT$ \cite{Strominger:2001pn,Witten:2001kn,Maldacena:2002vr}
states that quantum gravity in de Sitter space is dual to a Euclidean
CFT living on the future boundary $\cI^+$.  More specifically, the
CFT partition function with specified sources $\phi_{i0}(\vx)$ coupled
to operators $\cO_i$ is identified \cite{Maldacena:2002vr} with the 
bulk Hartle-Hawking ``wavefunction of the universe'' as a functional 
of the boundary values of the fields dual to $\cO_i$ given by 
$\phi_{i0}(\vx)$. In a semiclassical approximation, this becomes\ 
$Z_{CFT} = \Psi[\phi_{i0}(\vx)] \sim e^{iS_{cl}[\phi_{i0}]}$\
where we need to impose regularity conditions on the past cosmological
horizon $\tau\ra -\infty$:\ \eg\ scalar modes satisfy
$\phi_k(\tau)\sim e^{ik\tau}$, which are Hartle-Hawking (or
Bunch-Davies) initial conditions. Operationally, certain $dS/CFT$
observables can be obtained by analytic continuation (\ref{AdStodS})
from $AdS$ (see \eg\ \cite{Maldacena:2002vr}, as well as
\cite{Harlow:2011ke}). The Bunch-Davies initial condition itself can
be thought of as analytic continuation of regularity in the $AdS$
interior. Dual CFT correlation functions can be obtained from the 
dictionary $Z_{CFT}=\Psi$. For instance (from \cite{Maldacena:2002vr}) 
in a semiclassical approximation $\Psi\sim e^{iS_{cl}}$, a massless 
scalar in $dS_4$ has a mode solution
$\phi=\phi^0_{{\vec k}} {(1-ik\tau) e^{ik\tau}\over (1-ik\tau_c)e^{ik\tau_c}}$ 
with Bunch-Davies initial conditions $\phi\sim e^{ik\tau}$ at early 
times ($|\tau|\ra\infty$) and $\tau_c$ a late-time cutoff. The 
classical action evaluated on this solution is\ \
$iS_{dS_4} \sim {R_{dS}^2\over G_4} \int d^3k (i{k^3\over\tau_c} - k^3 
+ \ldots) \phi^0_{-{\vec k}} \phi^0_{{\vec k}}$~, the divergent terms 
being oscillatory (pure imaginary). Appropriate graviton modes can 
be approximated as massless scalars so that dual CFT energy-momentum 
tensor $\langle TT \rangle$ correlators can be read off as\ 
$\langle T({\vec k})T({\vec k}')\rangle_{dS_4} = 
{\delta^2 Z_{CFT}\over\delta\phi^0_{{\vec k}}\delta\phi^0_{{\vec k}'}}|_{\phi^0=0} 
\sim (-{R_{dS}^2\over G_4}) k^3 \delta({\vec k}+{\vec k}')$. The 
central charge coefficient here is ${\cal C}_4\sim -{R_{dS}^2\over G_4}$
essentially an analytic continuation from Euclidean $AdS_4$. 
In $dS_5$ \cite{Maldacena:2002vr}, we have\ 
$iS_{dS_5} \sim i{R_{dS}^3\over G_5} \int d^4k \phi^0_{-{\vec k}}\phi^0_{{\vec k}} 
(\ldots + k^4\log (-\tau_c k) + \ldots)$ where we have 
only exhibited the nonlocal term which contributes to the 2-point 
function: this gives 
$\langle T({\vec k})T({\vec k}')\rangle_{dS_5} \sim (i{R_{dS}^3\over G_5})
k^4\log k\ \delta({\vec k}+{\vec k}')$, with central charge coefficient 
${\cal C}_5\sim i{R_{dS}^3\over G_5}$. More generally we have 
${\cal C}_d\sim i^{1-d} {R_{dS}^{d-1}\over G_{d+1}}$\ (which is 
essentially an analytic continuation from $EAdS_{d+1}$).

From this $dS/CFT$ point of view, one might expect any leading 
divergence in the CFT entanglement entropy (assuming it exists) to be 
of the form ${\cal C}_d {V_{d-2}\over \epsilon^{d-2}}$ which is in general 
complex-valued: thus it is natural to ask if there are additional 
(perhaps complex\footnote{Recall that complex geodesics 
appeared in \cite{Fidkowski:2003nf} in the context of the black hole 
interior. Complex extremal surfaces have recently appeared in a different 
context \cite{Fischetti:2014zja}, as well as \cite{Fischetti:2014uxa}.}) 
extrema of the area functional that should be considered in 
de Sitter space, with possible $dS/CFT$ interpretations.
With a view to considering spacelike surfaces with ${\dot x}^2>1$, 
let us take, instead of (\ref{RTdS01}), the $dS_{d+1}$ area functional 
on a $w=const$ slice as
\be\label{SdS-EEdS}
S_{dS} = {R_{dS}^{d-1} V_{d-2}\over 4G_{d+1}} 
\int {d\tau\over\tau^{d-1}} \sqrt{\Big({dx\over d\tau}\Big)^2-1\ }\ ,
\qquad\qquad {{\dot x}\over\sqrt{{\dot x}^2-1}} = A\tau^{d-1}\ ,
\ee
where we are considering strip subsystems with width along $x$.
The second expression above is the conserved quantity obtained in 
the extremization. This is essentially the same as (\ref{RTdS02}), 
but with $B^2=-A^2<0$, and $A$ being real. One might ask if this can 
be interpreted as a real surface\ 
${\dot x}^2 = {A^2\tau^{2d-2}\over A^2\tau^{2d-2}-1}$. 
However we need to require that the surface reaches the boundary
$\tau\ra 0$ from where it drops down (inward): near $\tau\ra 0$, we
find\ ${\dot x}^2 \sim -A^2\tau^{2d-2}$, so that 
\be\label{xdotCmplx}
{\dot x}^2 = {-A^2\tau^{2d-2}\over 1-A^2\tau^{2d-2}}\ ,
\ee
this being a complex surface in some sense.

Let us focus now on $dS_4/CFT_3$ for concreteness, to understand this
better. The extremal surface near $\tau\ra 0$ in this case is\
$x(\tau) \sim \pm iA\tau^3 + x(0)$.  We want $x(\tau)$ to be
real-valued since it parametrizes a space direction in the dual
$CFT_3$: this requires that $\tau$ takes imaginary values. In more
detail, near $\tau\ra 0$, we have $x\ra \pm {l\over 2}$ and the two
ends of the surface are parametrized as\ 
$x_L(\tau)\sim -{l\over 2} + iA\tau^3$ and 
$x_R(\tau)\sim {l\over 2} - iA\tau^3$. For $x_{L,R}$ to be 
real-valued with $A$ real, we must have pure imaginary $\tau=iT$ 
with $T$ real, giving\ $x_L \sim -{l\over 2} + AT^3 \sim -x_R$:\ \ 
as $T$ increases, $x_L$ increases from $-{l\over 2}$ and $x_R$ 
decreases from ${l\over 2}$. The global structure of the surface 
shows a ``turning point'' at $\tau_*^4 A^2 = 1$, where 
${\dot x}^2\ra\infty$, very similar to the situation in $AdS$. From 
the point of view of the discussion in the previous subsection, the 
two half-extremal-surfaces $x_L, x_R$ in this case join smoothly 
at the turning point $\tau_*$ as in $AdS$, with 
$x_L(\tau_*)=0=x_R(\tau_*)$ and ${\dot x}_L, {\dot x}_R$ matching. 
This gives the width
\be
{\Delta x\over 2} = {l\over 2} =\ 
i\int_0^{\tau_*} {A\tau^{2}\ d\tau\over\sqrt{1-A^2\tau^{4}}}\ 
\xrightarrow{\ \tau=iT\ }\ \int_0^{T_*} {i.i^3.AT^2 dT\over\sqrt{1-A^2T^4}}
=\ \# T_*\quad \Rightarrow\quad \tau_*\sim il\ .
\ee
The reality of $\Delta x=l$ with $A$ real again suggests that we 
parametrize the $\tau$-integral over the path $\tau=iT$ in a complex 
$\tau$-plane\footnote{Strictly speaking, this may be too restrictive.
We have required $x(\tau)$ for all $\tau$ be real-valued: this means 
that each point on the surface directly maps to a corresponding 
real-valued spatial location within the strip in the dual $CFT_3$. 
One might instead think that one need only require the boundary value 
$x(0)$ be real, which would not restrict the $\tau$-path. This would 
suggest more general complex extremal surfaces defined over complex 
$\tau$-space, with the width $\Delta x$ required to be real-valued. 
See \eg\ \cite{Fischetti:2014uxa} for some discussions along these 
lines: I thank S. Fischetti for a discussion on this point.}: we 
have then rescaled $T$ using $A$ to make the integration 
variable dimensionless\ (and $\# = \int_0^1 {y^2 dy\over\sqrt{1-y^4}} 
= \sqrt{\pi} {\Gamma({3\over 4})\over\Gamma({1\over 4})}$ ).\ 
The turning point here is\ $\tau_*={i\over\sqrt{A}}$.
The integral can be parametrized in terms of hypergeometric functions 
${}_2F_1$.\ The extremal surface $x(\tau)$ with $\tau$ imaginary does 
not correspond to any real bulk subregion in $dS_4$ enclosed by the 
surface, but really lives in some auxiliary space. In a sense, the 
structure here is very much like analytic continuation of the $AdS_4$ 
expressions a la Ryu-Takayanagi: we will discuss this more below. 
From that point of view, since the analytic continuation 
(\ref{AdStodS}) faithfully maps $AdS_4\leftrightarrow dS_4$, this 
is a faithful map from the subsystem to the auxiliary bulk 
subregion.\ The area now becomes 
\be\label{EEdS4-0}
S_{dS}\ =\ 2{R_{dS}^{2}\over 4G_{4}} V_{1} \int_{\tau_{UV}}^{\tau_*} 
{d\tau\over\tau^{2}} {1\over\sqrt{A^2\tau^{4}-1}}\ =\ 
-i {R_{dS}^{2}\over 4G_4} V_{1} \int_{\tau_{UV}}^{\tau_*} 
{d\tau\over\tau^{2}} {2\over \sqrt{1-A^2\tau^{4}}}\ .
\ee
In principle, we could assign $\pm i$ in the second expression,
as a choice of the branch of the square root: the choice of the 
minus sign leads to an appropriate coefficient as we see below.
The integral itself is just as in $AdS_4$, giving
\be\label{EEdS4}
S_{dS}\ =\ -i {R_{dS}^{2}\over 2G_{4}} V_{1} 
\Big({1\over\tau_{UV}} - c_3 {1\over \tau_*}\Big)\ 
=\ - {R_{dS}^{2}\over 2G_{4}} V_{1} 
 \Big({1\over\epsilon} - c_3 {1\over l}\Big)\
\sim\ \ {\cal C}\  V_{1} \Big({1\over\epsilon} - c_3 {1\over l}\Big)\ ,
\ee
where $c_3=2\pi \big({\Gamma({3\over 4})\over\Gamma({1\over 4})}\big)^2$ 
is a constant as in $AdS$, stemming the finite cutoff-independent 
part of the integral. Note that here we have used the relation 
$\tau_{UV}=i\epsilon$ for the ultraviolet cutoff in the dual 
Euclidean field theory, suggested by previous 
investigations\footnote{See \eg\ \cite{Maldacena:2002vr,Harlow:2011ke,
Anninos:2011ui,Das:2013qea}, which discuss this (in some cases 
implicitly). Heuristically, we expect that evolution in the bulk 
direction is encoded by renormalization group flow in the dual 
field theory: see \eg\ \cite{Akhmedov:1998vf,Alvarez:1998wr,
Freedman:1999gp,de Boer:1999xf,de Haro:2000xn,Skenderis:2002wp} 
and more recently \eg\ \cite{Heemskerk:2010hk,Faulkner:2010jy} 
for discussions on this in the $AdS$ context. In the present 
$dS$ case, the bulk description is time evolution\ 
$i{\delta \Psi\over \delta \tau} = {\cal H}\Psi$, 
with ${\cal H}$ being an evolution operator. Through $dS/CFT$, this 
becomes\ $i{\delta Z_{CFT}\over \delta \tau} = {\cal H} Z_{CFT}$. 
Heuristically this maps to a renormalization group equation schematically 
of the form\ ${\delta Z_{CFT}\over \delta \epsilon} = {\cal O} Z_{CFT}$, 
if \ $\tau_{UV}=i\epsilon$, with ${\cal O}$ an appropriate 
operator generating RG flow. If we view $\epsilon$ as a floating 
RG parameter, this again suggests the path $\tau=iT$ in complex 
$\tau$-space for a $dS/CFT$ interpretation.} in $dS/CFT$.
Also we have rewritten the last expression in (\ref{EEdS4}) in 
terms of the $dS_4/CFT_3$ central charge\ 
${\cal C}_3\sim -{R_{dS}^2\over G_4}$ appearing in the 
$\langle TT \rangle$ correlators in \cite{Maldacena:2002vr}, reviewed 
above. $S_{dS}$ in (\ref{EEdS4}) is real-valued and bears structural 
resemblance to entanglement entropy in a dual CFT$_3$ with central 
charge ${\cal C}_3\sim -{R_{dS}^2\over G_4} < 0$. The first term 
resembles an area law divergence \cite{AreaLaw1,AreaLaw2},
proportional to the area of the interface between the subregion and
the environment, in units of the ultraviolet cutoff. It is also
proportional to the central charge which represents the number of
degrees of freedom in the dual CFT: in this case, ${\cal C}< 0$
reflecting the fact that the CFT is non-unitary. The second term is a
finite cutoff independent piece.  Whether the expression (\ref{EEdS4})
physically is holographic entanglement entropy in $dS_4/CFT_3$ is 
less clear from this bulk extremal surface analysis.

In some sense, $-S_{dS}$ appears to resemble entanglement entropy in
$AdS/CFT$, sharing various features including subadditivity. For
instance, the quantity\ $I_{dS}[A,B] = S_{dS}[A] + S_{dS}[B] -
S_{dS}[A\cup B]$ for two disjoint subsystems $A, B$, exhibits various
properties of holographic mutual information in $AdS$ including an
analog of the disentangling transition in the classical gravity
approximation \cite{Headrick:2010zt}, but with some crucial
differences. For strip subregions that are sufficiently nearby but
disjoint, $I_{dS}[A,B]$ is nonzero: \eg\ using (\ref{EEdS4}) for a 
single strip, we obtain for two parallel strips of equal width $l$ 
and separation $x$,
\be
I_{dS}[A,B] = S_{dS}[A] + S_{dS}[B] - S_{dS}[A\cup B] 
\sim\ - {R_{dS}^{2}\over G_{4}} c_3 V_1 \Big(-{2\over l} + {1\over x} + 
{1\over 2l + x} \Big) .
\ee
$S_{dS}[A\cup B]$ arises from the area of the connected surface
between $A, B$ as\ $S(2l+x)+S(x)$. This is similar to the structure 
of holographic mutual information for strips in $AdS_4$, \eg\ the 
UV divergent pieces cancel, with a cutoff-independent divergence 
${\cal C} {V_1\over x}$ as the subregions collide. The striking
difference is that $I_{dS}[A,B]\leq 0$, rather than positive definite,
following from the fact that ${\cal C} = -{R_{dS}^{2}\over G_{4}}
<0$. Thus $I_{dS}[A,B]$ is large and negative when the subregions are
nearby, then increases as the separation ${x\over l}$ increases, and
eventually approaches zero as\ ${x\over l} \ra {\sqrt{5}-1\over 2}\sim
0.62$. Beyond this critical value, $I_{dS}[A,B]$ vanishes identically
and the two subregions are disconnected. This disentangling transition
in the classical gravity approximation arises from the transition
between the connected and disconnected surface for $A\cup B$. What we
are seeing is that $S_{dS}[A]+S_{dS}[B]\leq S_{dS}[A\cup B]$, \ie\
$-S_{dS}$ satisfies strong subadditivity for disjoint parallel strips
$A, B$.  By comparison, using the real lightcone wedge surfaces in the
previous subsection, we see that disjoint boundary subregions are
always causally disconnected and thus uncorrelated for any nonzero
separation. Correlation functions are nonzero: the disentangling
transition above is in the classical gravity approximation, and we
expect subleading terms in a large-N-like expansion of $I_{dS}[A,B]$ 
(see \eg\ \cite{Faulkner:2013ana} in the $AdS$ context).

We now discuss $dS_{d+1}$ for even $d$ (in particular $dS_3, dS_5$)
where the nature of these extremal surfaces is different. We would
like to retain the relation\ $\tau_{UV}=i\epsilon$\ as following quite
generally in $dS/CFT$ from time evolution mapping to renormalization
group flow. This suggests we parametrize the bulk time parameter
$\tau$ along a complex path $\tau=iT$ as in $dS_4$. However now 
with $A^2>0$ the surface (\ref{xdotCmplx}) near $\tau\ra 0$ gives\ 
${\dot x} \sim \pm iA\tau^{d-1}$, \ie\ $x(\tau)\sim \pm iA\tau^d$.\
Thus $x$ cannot be made real-valued for any even $d$ in this manner. 
A way out is to take the parameter $A^2\ra -A^2$: the surface 
equation now is the same as (\ref{RTdS02}) but with the bulk time 
parametrized as $\tau=iT$. The expressions (\ref{SdS-EEdS}), 
(\ref{xdotCmplx}) then give 
\bea\label{EEdS-0}
{\dot x}^2 = {A^2\tau^{2d-2}\over 1+A^2\tau^{2d-2}}\ ,\quad
&& 
S_{dS}\ =\ -i {R_{dS}^{d-1}\over 4G_{d+1}} V_{d-2} \int_{\tau_{UV}}^{\tau_*} 
{d\tau\over\tau^{d-1}} {2\over \sqrt{1+A^2\tau^{2d-2}}}\ \nonumber\\
&& \qquad \xrightarrow{\tau=iT }\ 
i^{1-d} {R_{dS}^{d-1}\over 2G_{d+1}} V_{d-2} \int_{\epsilon}^{T_*} 
{dT\over T^{d-1}} {1 \over \sqrt{1+(-1)^{d-1} A^2 T^{2d-2}}}\ .\ \ 
\eea
For even $d$, the $(-1)^{d-1}$ gives rise to a ``turning point'' at 
$T_*^{2d-2} A^2 = 1$: the width now scales as $l\sim T_*\sim -i\tau_*$.
The integral is as in $AdS$, giving
\be\label{EEdS-1}
S_{dS}\ \sim\ i^{1-d} {R_{dS}^{d-1}\over 2G_{d+1}} V_{d-2} 
 \Big({1\over\epsilon^{d-2}} - c_d {1\over l^{d-2}}\Big)\ .
\ee
The leading divergence\ $S_{dS}^{div}\sim i^{1-d} {R_{dS}^{d-1}\over 2G_{d+1}} 
{V_{d-2}\over\epsilon^{d-2}}$ resembles an area law: the central 
charges\ ${\cal C}_d\sim i^{1-d} {R_{dS}^{d-1}\over G_{d+1}}$ here 
resemble those in the $\langle TT \rangle$ correlators in
\cite{Maldacena:2002vr} reviewed above. This leading behaviour appears 
independent of the shape of the subregion, expanding (\ref{SdS-EEdS}) 
and assuming that ${\dot x}$ is small near the boundary $\tau_{UV}$. 
Unlike $dS_4$, note that $S_{dS}$ in $dS_{d+1}$ with even $d$ is not 
real-valued, in particular for $dS_3, dS_5$. For instance, in 
$dS_3$, we obtain from (\ref{EEdS-0}) 
\be
\tau=iT\ ,\quad x(\tau)\sim \pm {1\over A} \sqrt{1+A^2\tau^2}\ ,\qquad 
S_{dS} \sim\ -i{R_{dS}\over G_3} 
\log {\tau_*\over\tau_{UV}} = -i{R_{dS}\over G_3} \log {l\over\epsilon}\ .
\ee
Note that $x(\tau)$ appears real, although the parametrization 
is $\tau=iT$.

It is interesting to recall the Ryu-Takayanagi expression for
entanglement entropy for an (infinitely long) strip-shaped subsystem
with width along the $x$-direction, given as the area of the
corresponding minimal surface in the bulk $AdS_{d+1}$ geometry (with
radius $R$),
\be\label{RTAdS}
S_{AdS}[R,x(r),r] = {R^{d-1}\over 4G_{d+1}} V_{d-2} 
\int {dr\over r^{d-1}}\sqrt{1+\Big({dx\over dr}\Big)^2}\ ,
\qquad\ \ (x')^2 = {A^2 r^{2d-2}\over 1 - A^2 r^{2d-2}}\ ,
\ee
where the conserved quantity $A$ in the extremization is related to 
the turning point as $r_*^{d-1}={1\over A}$.
Noting that $dS_{d+1}$ in Poincare slicing (\ref{dSPoinc-wslice}) 
is just the analytic continuation of the corresponding $t=constant$ 
spatial slice in $AdS_{d+1}$, obtained by (\ref{AdStodS}), \ie\
$r\ra -i\tau,\ t\ra -iw,\ R\ra -iR_{dS}$, let us carry out this 
analytic continuation on the Ryu-Takayanagi expression. Indeed we 
see that $S_{dS}$ in (\ref{RTdS01}) appears very 
much like the analytic continuation of $S_{AdS}[x(r),r]$,\ with the 
various factors of $i$ conspiring to leave a single $i$ behind, \ie\
\be\label{RTAdStodS1}
S_{AdS}[R,x(r),r]\ \ra\ -i\ {R_{dS}^{d-1}\over 4G_{d+1}} V_{d-2} 
\int {d\tau\over \tau^{d-1}}\sqrt{1-\Big({dx\over d\tau}\Big)^2}\ 
=\ S_{dS}[R_{dS},x(\tau),\tau]\ .
\ee
On the analytic continuation of the extremization itself, we obtain
\bea\label{RTAdStodS2}
&& -{\dot x}^2 = {(-i)^{2d-2} A^2\tau^{2d-2}\over
1-(-i)^{2d-2}A^2\tau^{2d-2}}\ , \qquad 
\ie\ \qquad  {\dot x}^2 =\ {- (-1)^{d-1} A^2\tau^{2d-2}\over
1-(-1)^{d-1}A^2\tau^{2d-2}}\ , \nonumber\\
&& S_{dS} = -i {R_{dS}^{d-1}\over 4G_{d+1}} V_{d-2} 
\int {d\tau\over\tau^{d-1}} 
{1\over\sqrt{1-(-1)^{d-1}A^2\tau^{2d-2}}}\ .
\eea
This expression corroborates the minus sign in (\ref{EEdS-0}) and 
(\ref{EEdS4-0}), (\ref{EEdS4}). The analytic continuation essentially 
recovers our earlier calculations in $dS_4$ and $dS_{d+1}$ for even $d$.
For instance, in $dS_5$ (\ie\ $d=4$), 
we obtain
\be\label{eeds5}
{\dot x}^2 =\ { A^2\tau^{6}\over 1+A^2\tau^{6}}\ , \qquad\qquad 
S_{dS} = -i {R_{dS}^{3}\over 4G_{5}} V_{2} \int {d\tau\over\tau^{3}} 
{1\over\sqrt{1+A^2\tau^{6}}}\ .
\ee
With real $A$, this is as such a real extremal surface as in the 
previous subsection: taking $A$ large minimises the area and we 
obtain the null surfaces earlier with vanishing area representing 
the past lightcone wedge of the subregion. However parametrizing 
as $\tau=iT$, there is a turning point at\ $\tau_*={i\over A^{1/3}}$, 
and a corresponding complex surface and corresponding area given by 
(\ref{EEdS-0}). The area in (\ref{eeds5}) then becomes\
$S_{dS}\sim i {R_{dS}^{3}\over 4G_{5}} V_{2} 
({1\over\epsilon^2}-c_4 {1\over l^2})$. The extra $i$ can be thought 
of as arising from the odd powers of $R_{dS}$ under the analytic
continuation from $AdS_5$. Thus interestingly for even $d$
(in particular, $dS_3$ and $dS_5$), the expression $S_{dS}$ obtained
by analytic continuation of the Ryu-Takayanagi entanglement
prescription leads to complex surfaces with corresponding area 
$S_{dS}$ pure imaginary.

To summarize, we have studied bulk de Sitter codim-2 extremal
surfaces.  Real extremal surfaces are the boundaries of the past
lightcone wedges of the boundary subregions, with vanishing area.
Codim-2 complex extremal surfaces have area exhibiting structural
resemblance with entanglement entropy in a dual CFT$_d$ with ${\cal
  C}_d\sim i^{1-d} {R_{dS}^{d-1}\over G_{d+1}}$ central
charge\footnote{Note that codim-1 surfaces with area functional\
  $S\sim {R_{dS}^d V_{d-1}\over l_P^d} \int {d\tau\over\tau^d}
  \sqrt{{\dot x}^2 - 1}$ upon extremization exhibit area whose leading
  divergence does not reflect the CFT central charge.} matching those
appearing in the $\langle TT\rangle$ correlators obtained from the
wavefunction of the universe \cite{Maldacena:2002vr}. In $dS_4/CFT_3$,
the area is real-valued and negative: in this sense, these complex
surfaces have lower area, suggesting that they are the preferred
minimal surfaces.  Our calculations here have been done for a strip
subregion but it would appear that generalizations to other subregion
shapes will exhibit similar features. For instance, the spherical
subregion extremal surface presumably exhibits a logarithmic term with
interesting coefficient\ (this universal coefficient in the
logarithmic term has been studied recently in \cite{Narayan:2015oka},
exhibiting agreement with the corresponding coefficient in the
logarithmically divergent term in the wavefunction of the universe 
in the classical approximation).

It is worth noting that this analysis of bulk extremal surfaces is
different from studies of entanglement entropy of bulk fields in
de Sitter space \eg\ \cite{Maldacena:2012xp,Kanno:2014lma,
Iizuka:2014rua,Fischler:2013fba}.

\section{Extremal surfaces in the $dS$ black brane}

We now study extremal surfaces in the 
asymptotically $dS$ spacetime studied in \cite{Das:2013mfa}, \ie\
\be\label{dSbwCmplx}
ds^2 = {R_{dS}^2\over\tau^2} \Big( -{d\tau^2\over 1+\al\tau_0^d\tau^d}
+ (1+\al\tau_0^d\tau^d) dw^2 + \sum_{i=1}^{d-1} dx_i^2 \Big)\ ,
\ee
with $\al$ a complex phase and $\tau_0$ is some real parameter of 
dimension length inverse.
An analog of regularity in the interior for an asymptotically $AdS$ 
solution is obtained here by a Wick rotation $\tau=il$ and demanding 
that the resulting spacetime (thought of as a saddle point in a path 
integral) in the interior approaches flat Euclidean space in the 
$(l,w)$-plane with no conical singularity. This makes 
the $w$-coordinate angular with fixed periodicity (and $l$ is a 
radial coordinate), giving\  
$\al = -(-i)^d ,\ l\geq \tau_0 ,\ w\simeq w + {4\pi\over (d-1)\tau_0}$.
Thus the spacetime (\ref{dSbwCmplx}) is a complex metric which 
satisfies Einstein's equation with a positive cosmological constant\ \ 
$R_{MN}={d\over R_{dS}^2}g_{MN}~, \ \ \Lambda={d(d-1)\over 2R_{dS}^2}$ .
This resulting metric satisfying regularity is equivalent to one 
obtained by analytically continuing the Euclidean $AdS$ black brane 
\be\label{EAdSbb}
ds^2 = {R_{AdS}^2\over r^2} \Big( {dr^2\over 1-r_0^dr^d} 
+ (1-r_0^dr^d) d\theta^2 + \sum_{i=1}^{d-1}dx_i^2 \Big)  ,
\ee
where $\theta \sim \theta + \frac{4\pi}{(d-1)r_0}$~, to the 
asymptotically de Sitter spacetime (\ref{dSbwCmplx}) using 
(\ref{AdStodS}) and we identify $r_0\equiv\tau_0$, giving the phase 
${-1\over (-i)^d}$. The regularity criterion is simply the analog of 
regularity of the $EAdS$ black brane. The condition $l\geq \tau_0$ is 
equivalent to the radial coordinate having the range $r\geq r_0$.\ 
We see that ``normalizable'' metric pieces are turned on in 
(\ref{dSbwCmplx}). We then expect a nonzero expectation value for 
the energy-momentum tensor here, as in the $AdS$ context 
\cite{Balasubramanian:1999re,Myers:1999psa,de Haro:2000xn,Skenderis:2002wp}.
In the present case \cite{Das:2013mfa}, we have\ 
$T_{ij} = {2\over\sqrt{h}} {\delta Z_{CFT} \over\delta h^{ij}} 
= {2\over\sqrt{h}} {\delta \Psi\over\delta h^{ij}} 
\propto i{R_{dS}^{d-1}\over G_{d+1}} g^{(d)}_{ij}$,\ 
where $g^{(d)}_{ij}$ is the coefficient of the 
normalizable $\tau^{d-2}$ term in the Fefferman-Graham expansion of 
the metric (\ref{dSbwCmplx}). This definition of $T_{ij}$ is natural 
for a CFT with partition function $Z_{CFT}$, equated with $\Psi$: 
thus, most notably, the $i$ arising from $\Psi$, the wavefunction 
of the universe, implies that the energy-momentum tensor is real 
only if $g^{(d)}_{ij}$ is pure imaginary.
In effect, this $dS/CFT$ energy-momentum tensor can be thought of as 
the analytic continuation of the $EAdS$ one. The spacetime 
(\ref{dSbwCmplx}) for $dS_4/CFT_3$ gives real $T_{ij}$, with\ 
$T_{ww}=-{R_{dS}^2\over G_{4}}\tau_0^3$\ with $T_{ww}+(d-1)T_{ii}=0$.

The $w$-coordinate is naturally interpreted as Euclidean time from 
the structure of the energy-momentum tensor: so let us now consider 
a strip subregion on a $w=const$ surface in (\ref{dSbwCmplx}). The 
area functional (in Planck units) of a bulk surface bounding this 
strip and dipping inwards is
\be
S_{dS}\ =\ -i{R_{dS}^{d-1}\over G_{d+1}} V_{d-2} 
\int {d\tau\over \tau^{d-1}}\sqrt{{1\over 1+\al\tau_0^d\tau^d} 
-\Big({dx\over d\tau}\Big)^2}\ ,
\ee
defined so that for $\tau_0=0$, this reduces to our de Sitter 
discussion in sec.~2.2.\ 
For the $dS_4$ brane (\ie\ $d=3$), we obtain for the extremization,
\be
{\Delta x\over 2} = \int_{0}^{\tau_*} {iA\tau^{2}\ d\tau\over\sqrt{
\big(1-i\tau_0^3\tau^3\big) \big(1-A^2\tau^{4}\big)}} ,\quad
S = -i{V_{1}R_{dS}^{2}\over 4G_{4}} \int_{\tau_{UV}}^{\tau_*} 
{d\tau\over \tau^{2}} {2\over\sqrt{\big(1-i\tau_0^3\tau^3\big) 
\big(1-A^2\tau^{4}\big)}} .
\ee
Now for small width $l$, this is essentially similar to the previous 
discussion on pure $dS_4$ and we have\ 
${i\Delta x\over 2} \equiv {il\over 2} \sim \tau_* = {i\over\sqrt{A}}$, 
where $A$ is real. In particular, the width $\Delta x$ being 
real-valued suggests that $\tau$ parametrizes a complex path $\tau=iT$ 
with $T$ real. As $l$ increases however, the other denominator 
approaches a zero also, with $\tau\ra {i\over\tau_0}$. In this limit, 
we thus have\ $\tau_* \ra {i\over\sqrt{A}} \sim {i\over\tau_0}$ and 
large $l\sim -i\tau_*$, obtained from the double zero as\
\be
{\Delta x\over 2} ={l\over 2} \sim 
\int_{0}^{\tau_*} {iA\tau_*^{2}\ d\tau\over\sqrt{
(1-i\tau_0^3\tau^3) (1-A^2\tau^{4})}} \xrightarrow{\tau=iT} 
\int_0^{T_*} {i.i^3AT_*^2\ dT\over \sqrt{(1-\tau_0^3T^3)(1-A^2T^4)}}\ .
\ee
Note that reality of the width $\Delta x$ implies now that the range 
of $T$ is restricted as $T\leq {1\over\tau_0}$ \ie\ asymptotically 
$\tau\ra {i\over\tau_0}$. This is similar to the fact that in the 
$AdS$ black brane, static minimal surfaces in the IR limit (large 
subsystem width) wrap the horizon but do not penetrate beyond.

Now the area integral exhibits a cutoff-independent piece
which can be estimated from the contribution in the deep interior 
where $\tau\ra\tau_*$: the contribution to the integral near the 
double zero thus scales as $i\Delta x$ giving
\be
S^{fin}\ \sim\ -i{V_{1}R_{dS}^{2}\over G_{4}} {1\over\tau_*^2} (il)\ 
=\  - {R_{dS}^{2}\over G_{4}} \tau_0^2 V_{1}l\ \equiv\ 
{\cal C} T_0^2 V_l l\ ,
\ee
which resembles an extensive thermal entropy in a 3-dim CFT with 
central charge ${\cal C}\sim -{R_{dS}^{2}\over G_{4}}$ at 
temperature $T_0\equiv \tau_0$.\ Note that $S^{fin}<0$. In fact $S^{fin}$ 
is the analytic continuation (\ref{AdStodS}) of the horizon 
entropy ${R_{AdS}^{2}\over G_{4}} \tau_0^2 V_{1}l$ of the $EAdS_4$ 
black brane (\ref{EAdSbb}), which can be obtained as the horizon 
contribution from the partition function in the classical approximation.

We recall that the entanglement entropy area functional for the 
$AdS_{d+1}$ black brane from the Ryu-Takayanagi prescription is\
$S = {V_{d-2}R^{d-1}\over 4G_{d+1}} \int {dr\over r^{d-1}} 
\sqrt{(\del_rx)^2+{1\over 1-r_0^dr^d}}$,\ giving
\be\label{EEadsBB}
(x')^2 = {A^2r^{2d-2}\over (1-r_0^dr^d) (1-A^2r^{2d-2})} ,
\qquad S = {V_{d-2}R^{d-1}\over 4G_{d+1}} \int_\epsilon^{r_*} 
{dr\over r^{d-1}} {2\over\sqrt{(1-r_0^dr^d) (1 - A^2r^{2d-2})}} .
\ee
Under the analytic continuation, we obtain
\bea
&& {\dot x}^2 = {-(-1)^{d-1}A^2\tau^{2d-2}\over
\big(1-(-i)^d\tau_0^d\tau^d\big) \big(1-(-1)^{d-1}A^2\tau^{2d-2}\big)}\ ,
\nonumber\\ 
&& S = {V_{d-2}R_{dS}^{d-1}\over 4G_{d+1}} \int_{\tau_{UV}}^{\tau_*} 
{-id\tau\over \tau^{d-1}} {2\over\sqrt{\big(1-(-i)^d\tau_0^d\tau^d\big) 
\big(1-(-1)^{d-1}A^2\tau^{2d-2}\big)}}\ .
\eea
For generic dimension $d$, we see that $S$ is not real, as in 
the earlier discussion with $\tau_0=0$.

\subsection{The de Sitter bluewall}

\begin{figure}[h] 
\hspace{1pc} \includegraphics[width=14pc]{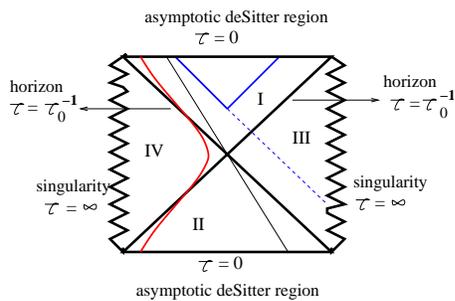} \hspace{3pc}
\begin{minipage}[b]{20pc}
\caption{{\label{dsbw-extsur}\footnotesize {de Sitter ``bluewall'' 
Penrose diagram and \newline
some extremal surfaces with at least one end anchored \newline 
at ${\cal I}^+$. The blue wedge is null, while the red timelike 
\newline surface goes from ${\cal I}^+$ to ${\cal I}^-$. \newline\newline
}}}
\end{minipage}
\end{figure}

We now explore metrics of the form (\ref{dSbwCmplx}), but with the 
parameter $\al=-1$ here\footnote{The metric (\ref{dSbwCmplx}) with 
$\al=+(-i)^d$ is similar to the $dS$ black brane, except with 
$T_{ij}$ of the opposite sign, while $\al=+1$ gives a real spacetime 
with a spacelike singularity at $\tau\ra\infty$.}, \ie\
\be\label{dSwh}
ds^2 = {R_{dS}^2\over\tau^2} \Big( -{d\tau^2\over 1-\tau_0^d\tau^d} 
+ (1-\tau_0^d\tau^d) dw^2 + dx_i^2 \Big) \equiv\ 
{R_{dS}^2\over\tau^2} \Big( -{d\tau^2\over f(\tau)} 
+ f(\tau) dw^2 + dx_i^2 \Big)\ .
\ee
The $w$-coordinate here has the range $-\infty\leq w\leq \infty$.
This spacetime \cite{Das:2013mfa} has a Penrose diagram shown in 
Figure~\ref{dsbw-extsur} which resembles that of the $AdS$ black 
brane rotated by ${\pi\over 2}$: there are two asymptotically 
$dS$ universes (for $\tau\lesssim {1\over\tau_0}$), 
and timelike singularities cloaked by Cauchy horizons at 
$\tau={1\over\tau_0}$, which ``cross'' at a bifurcation region. 
The Penrose diagram has many similarities with the interior of the
Reissner-Nordstrom black hole (or wormhole). Late time infalling
observers near the Cauchy horizon see incoming lightrays from early
times as highly blueshifted, essentially stemming from lightrays
``crowding'' near the Cauchy horizon, suggesting an instability.  It
is unclear if this spacetime has any interpretation in $dS/CFT$:
nevertheless, formally, one finds the energy-momentum $T_{ij}$ to 
be imaginary in $dS_4/CFT_3$, perhaps reflecting the blueshift 
instability here. Here we will simply look for bulk codim-2 extremal 
surfaces, lying either on a $w=const$ slice or an $x=const$ slice 
(from a bulk point of view alone, either could be taken as 
Euclidean time slices), restricting to real surfaces which may 
also be timelike.

The area functional for a surface in (\ref{dSwh}) bounding a subregion 
on a $x=const$ slice of ${\cal I}^+$, and wrapping the other 
$x_i\neq x$, is
\be
S\ =\ {R_{dS}^{d-1}\over 4G_{d+1}} V_{d-2} 
\int {d\tau\over \tau^{d-1}}\sqrt{{1\over f(\tau)} 
-f(\tau) \Big({dw\over d\tau}\Big)^2}\ ,\qquad f(\tau)=1-\tau_0^d\tau^d\ .
\ee
This does not correspond to any analytic continuation from the 
Ryu-Takayanagi formula for the $AdS_4$ black brane, so we analyse 
this directly focussing on the $dS_4$ bluewall. Along our earlier 
discussions in sec.~2, we find real extremal surfaces corresponding to
\be
{\dot w}^2 = {1\over (1-\tau_0^3\tau^3)^2} {B^2\tau^4\over 
1-\tau_0^3\tau^3 + B^2\tau^4}\ ,\qquad
S = {V_{1}R_{dS}^{2}\over 4G_{4}} \int {d\tau\over\tau^{d-1}} {1\over
\sqrt{1-\tau_0^3\tau^3 + B^2\tau^4}}\ ,
\ee
where the constant $B$ arises from a conserved quantity in the 
extremization. The first equation describing the surface can be 
rewritten as
\be\label{w'*2}
d\tau_* \equiv {d\tau\over 1-\tau_0^3\tau^3}\ ,\qquad\qquad
\Big({dw\over d\tau_*}\Big)^2 \equiv (w_*')^2 = {B^2\tau^4\over 
1-\tau_0^3\tau^3 + B^2\tau^4}\ ,
\ee
where we are using $\tau_*$ here for the ``tortoise'' coordinate 
in this bluewall geometry \cite{Das:2013mfa}, analogous to the 
Schwarzschild tortoise coordinate $r_*$. Parametrized thus, we see 
as in the $dS_4$ case that increasing $B$ decreases the area, as 
long as we restrict the surface to lie within the future asymptotic 
universe $I$, \ie\ $f(\tau)>0$. \ As $B^2\ra\infty$, these extremal 
surfaces become null with $(w_*')^2 = 1$, corresponding to the past 
lightcone wedges of the boundary subregion, and have vanishing area. 
Thus extremal surfaces 
for a given subregion at ${\cal I}^+$ can be constructed as in $dS_4$ 
(Figure~\ref{eedS}) by joining two half-extremal surfaces: this is 
the blue wedge in region $I$ in Figure~\ref{dsbw-extsur} (the 
half-surface when not cut off continues as a null surface through 
the Cauchy horizon into region $III$, represented by the dotted 
extension of the blue line). As the subregion grows in size, this 
blue wedge approaches and eventually wraps the future Cauchy horizons.

One might imagine that there are timelike surfaces which are not 
restricted to just region $I$ but instead start on ${\cal I}^+$ in 
$I$ and cross over to $II$ ending on the past boundary ${\cal I}^-$. 
These can be found with the parameter $B^2>0$ being finite. In this 
case, we see that $(w_*')^2\ra 0$ as $\tau\ra 0$ and $(w_*')^2\ra 1$ 
as $\tau\ra {1\over\tau_0}$ near the horizon in $I$. Now after the 
surface crosses the future Cauchy horizon, we have $f(\tau)<0$ in $IV$. 
Requiring that $(w_*')^2$ in (\ref{w'*2}) satisfies $(w_*')^2\geq 0$
corresponding to real surfaces, it is possible to see (\eg\ by
plotting as a function of $\tau_0\tau$) that the parameter $B^2$ is
bounded below by a critical value. There is a family of such surfaces: 
we will isolate one ``critical'' surface for a particular value of 
$B$, in what follows. Drawing analogies with the study of the phase 
transition found in \cite{Narayan:2012ks} (although the physical 
context there is different), we note that $(w_*')^2\ra\infty$ when 
the denominator in (\ref{w'*2}) approaches a double zero (with 
$y\equiv\tau_0\tau$), \ie\
\be
1-y_c^3+{B^2\over\tau_0^4} y_c^4 = 0\ ,\quad 
-3y_c^2+4{B^2\over\tau_0^4} y_c^3 = 0\ \ \ \ \Rightarrow\quad\ 
{B^2\over\tau_0^4} = {3\over 4. 4^{1/3}}\ ,\quad  y_c=4^{1/3}\ .
\ee
This corresponds to $\tau_c={4^{1/3}\over\tau_0}\sim {1.6\over\tau_0}$
which is just a little inside the Cauchy horizon in region $IV$.
Note that $(w_*')^2|_{\tau_c}\ra\infty$ here means this curve is
normal to the $w=const$ line here (these are straight spacelike lines
passing through the bifurcation point and hitting the singularity in
$IV$), or equivalently tangent to the $\tau=const$ curve at $\tau_c$ 
in $IV$. The corresponding surface from $\tau=0$ to $\tau=\tau_c$ can 
be drawn as a curve in the $(\tau,w)$-plane: it can be joined 
smoothly at $\tau_c$ with a corresponding curve from ${\cal I}^-$, 
resulting approximately in the red curve in Figure~\ref{dsbw-extsur}. 
This surface crosses the upper and lower Cauchy horizons at 
$\tau={1\over\tau_0}, w=+\infty$ and $\tau={1\over\tau_0}, w=-\infty$. 
The area of this surface has a leading divergence\ 
$S\sim {R_{dS}^2\over G_4} {V_1\over\tau_{UV}}$.  
Near the double zero, $\Delta w$ acquires a large contribution and 
we can estimate\ $S\sim {R_{dS}^2\over G_4} V_1 \Delta w$. 
This surface is vaguely reminiscent of the extremal surface in
\cite{Hartman:2013qma} which goes from one timelike boundary to the
other: since the $dS$ bluewall metric itself is related to the
AdS-Schwarzschild black brane by flipping minus signs, it is perhaps
not surprising that there exists a similar surface here (but
timelike), albeit with no obvious corresponding interpretation. In
light of ER=EPR \cite{Maldacena:2013xja}, it is amusing to speculate
that the subregion here corresponds to copies on both ${\cal I}^\pm$
possibly ``entangled'', in some sense, thinking of the bluewall
geometry as a ``timelike wormhole'' with the bifurcation region being
the Einstein Rosen bridge. Note however that strictly speaking, all
timelike geodesics go from ${\cal I}^-$ to ${\cal I}^+$  (unlike a 
shortcut in spacetime) either through the bifurcation region or 
through the Cauchy horizons, subject to the blueshift instability 
\cite{Das:2013mfa}.

With a $w=const$ slice, real extremal surfaces likewise have
\be
{\dot x}^2 = {B^2\tau^4 \over (1-\tau_0^3\tau^3)(1+B^2\tau^4)} ,\qquad
S = {V_{1}R_{dS}^{2}\over 4G_{4}} \int_{\tau_{UV}}^{T_0} 
{d\tau\over \tau^{2}} {2\over\sqrt{\big(1-\tau_0^3\tau^3\big) 
\big(1+B^2\tau^{4}\big)}} .
\ee
For $B^2\ra\infty$, these are again null extremal surfaces 
${\dot x}^2 = {1\over 1-\tau_0^3\tau^3}$ with vanishing area. 
These surfaces all lie on a $w=const$ slice (thin black straight 
line from ${\cal I}^+$ to ${\cal I}^-$ in Figure~\ref{dsbw-extsur}).

\section{Discussion}

We have considered extremal surfaces in bulk de Sitter space (in the
Poincare slicing) on constant boundary Euclidean time slices bounding
subregions at future timelike infinity, motivated by the
Ryu-Takayanagi prescription for entanglement entropy in $AdS/CFT$.
Stemming from certain crucial sign differences, we have seen real
extremal surfaces which are essentially the restrictions of the
boundaries of the past lightcone wedges of the subregion: these are
null surfaces with vanishing area. We have also seen complex extremal
surfaces which do not always have real-valued area: this has parallels
with analytically continuing from the Ryu-Takayanagi formula in
$AdS$. In $dS_4$, the area is real-valued and negative. The area has
structural resemblance with entanglement entropy in a dual $CFT_d$,
with the leading divergence of the form ${\cal C}_d
{V_{d-2}\over\epsilon^{d-2}}$ :\ the central charges\ ${\cal C}_d\sim
i^{1-d} {R_{dS}^{d-1}\over G_{d+1}}$ here resemble those in the CFT
energy-momentum tensor $\langle TT \rangle$ correlators in
\cite{Maldacena:2002vr} obtained in $dS/CFT$ using $Z_{CFT}=\Psi$ and
a semiclassical approximation $\Psi\sim e^{iS_{cl}}$ for the
wavefunction of the universe.  Alternatively, the strip entanglement
entropy of a nonunitary CFT with these central charges would be of
this form (assuming it exists), which the area of these complex
extremal surfaces reproduces. This appears distinct from bulk
entanglement entropy in de Sitter space (from a bulk density matrix
via the wavefunction $\Psi$).  We have also studied extremal surfaces
in the $dS$ black brane (where there is a finite cutoff-independent
extensive piece), and the related $dS$ bluewall spacetime.  It is
worth mentioning that there may exist other extrema of the area
functional: for instance, we have required that $x(\tau)$
parametrizing the strip width be real-valued, which suggests the path
$\tau=iT$ in complex $\tau$-space.  This appears consistent with
possible $dS/CFT$ interpretations and also corroborates with our
discussion of the $dS$ black brane. However this may be restrictive
and more general complex extremal surfaces may be relevant in complex
$\tau$-space (see \eg\ \cite{Fischetti:2014uxa}). It may be
interesting to understand if the analysis of \cite{Lewkowycz:2013nqa}
can be applied in this case to obtain insights into extremal surfaces.

While this analysis of bulk extremal surfaces could be regarded as
simply a study of certain kinds of probes of asymptotically de Sitter
spaces, it cannot pinpoint whether the corresponding area is expected
to have a physical interpretation as entanglement entropy in the dual
CFT, 
although the results do appear so, keeping in mind the central 
charges of the non-unitary CFTs in question. It is tempting to
study this in light of the higher spin $dS_4/CFT_3$ duality of
\cite{Anninos:2011ui}.  However the presence of massless higher spin
fields might suggest that extremal surfaces which are geometric
gravitational objects are not accurate (see \eg\
\cite{deBoer:2013vca,Ammon:2013hba} which study entanglement entropy
from Wilson lines in higher spin $AdS$ holography). Nevertheless it is
interesting to ask if these extremal surfaces have significance in
some approximation where the higher spin symmetry is not exact. In
this case, it would be interesting to explore the physical
interpretation here more directly from a Euclidean $CFT_3$ point of
view. One way to think about entanglement entropy in field theory
(lattice models) is in terms of the eigenvalues of a correlation
matrix and a corresponding von Neumann entropy (see \eg\
\cite{peschel} and more recently \cite{Herzog:2012bw}). In that
context, a simple model of a massless scalar field with wrong sign
kinetic terms might suggest that the correlation matrix squared $C^2$
is related to that for an ordinary massless scalar field by a minus
sign, so that $C$-eigenvalues $\lambda_k$ become $i\lambda_k$. Then
the associated von Neumann entropy is in general not real-valued: it
would be interesting to understand this better.

\vspace{4mm}

{\footnotesize \noindent {\bf Acknowledgements:}\ \
It is a pleasure to thank the participants of the ``Entanglement 
from Gravity'' Discussion meeting, ICTS, Bangalore, in particular
S. Banerjee, G. Mandal and R. Myers for helpful comments. I also thank
the organizers of this workshop and the Indian Strings Meeting 2014,
Puri, for hospitality as this was being completed. This work is
partially supported by a grant to CMI from Infosys Foundation.}


\end{document}